\documentclass[12pt]{iopart}
\usepackage{amssymb,epsfig,graphicx,multirow,adjustbox,multicol}

\begin{document}
\title{Antisite disorder and phase segregation in Mn$_{2}$NiSn}
\author{S V Malik $^1$, E T Dias $^1$, A K Nigam$^2$ and K R Priolkar$^1$\footnote[1]{Author to whom correspondence should be addressed}}
\address{$^1$School of Physical and Applied Sciences, Goa University, Goa, 403206, India.}
\address{$^2$Tata Institute of Fundamental Research, Dr. Homi Bhabha Road, Colaba, Mumbai, 400005, India.}
\ead{krp@unigoa.ac.in}
\date{\today}

\begin{abstract}
A systematic study of crystal structure, local structure, magnetic and transport properties in quenched and temper annealed Ni$_{2-x}$Mn$_{1+x}$Sn alloys indicate the formation of Mn$_3$Sn type structural defects caused by an antisite disorder between Mn and Sn occupying the Y and Z sublattices of X$_2$YZ Heusler structure. The antisite disorder is caused by the substitution of Ni by Mn at the X sites. On temper annealing, these defects segregate and phase separate into $L2_1$ Heusler and $D0_{19}$ Mn$_3$Sn type phases.
\end{abstract}

\maketitle

\section{Introduction}
Inverse Heusler alloys of the type Mn$_{2}$XZ crystallizing in cubic or tetragonal structures have the Mn atoms occupy both the octahedral (0,0,0) or (0.5, 0.5, 0.5) and the tetrahedral sites (0.25, 0.25, 0.25) or (0.75, 0.75, 0.75) with antiparallel spins of different magnitudes resulting in a ferrimagnetic order \cite{Graf.39.1-50,felser.46.668--699,Wollmann.90.214420,Wollmann.92.064417}. These inverse Heusler alloys display a high Curie temperature ($T_C$) and outstanding functionalities due to a strong interplay between the crystal structure and magnetic properties. Mn$_{2}$NiGa exhibits a shape memory effect with a magnetic field induced strain up to 4\%, a martensitic transformation temperature, $T_M$ = 270 K and a $T_C$ = 588 K raising a high prospect for achieving a magnetic shape memory alloy with improved properties \cite{Lui.87.262504}. Density functional theory (DFT) calculations indicate similar possibilities in Mn$_2$NiAl \cite{kadri.92.699-706}. Mn$_{2}$VAl behaves like a half-metallic ferromagnet, with the Mn moments associated with the two different crystallographic sites coupled parallel to one another \cite{Jiang.118.513-516}. Mn$_{2}$CoAl is reported as a spin gapless semiconductor with a very high magnetic ordering temperature of 720 K and also exhibits an anomalous Hall effect \cite{Ouardi.110.100401}. Topological Hall effect has been reported in Mn$_2$PtIn and Mn$_{2}$PtSn inverse Heusler alloys \cite{PhysRevB.102.014449, Li5039921}. Further, large exchange-bias effect, even under zero-field cooled conditions is displayed by Mn$_2$PtSn, Mn$_{2}$PtGa \cite{PhysRevB.102.014449, Li5039921,Nayak.110.127204}and Mn$_{2}$RhSn  alloys \cite{Meshcheriakova.113.087203}. Such tetragonally distorted doped inverse Heusler alloys also display novel noncollinear spin textures or antiskyrmions \cite{Nayak2017,acs.nanolett9b02973}. A high degree of spin polarization is reported in bulk as well as thin-film forms of Mn$_{2}$CoGa \cite{Winterlik.99.222510}. Mn$_{2}$FeGa behaves like an exchange spring, raising hopes of applications in magnetic storage technology \cite{Teuta.102.202402}.

Mn$_{2}$NiSn is one such inverse Heusler alloy that orders magnetically at $T_C$ = 530 K \cite{Luo32140634066} and has gained attention due to the predicted volume-conserving martensitic transformation in the magnetically ordered state leading to the observation of the shape memory effect \cite{Duan.386.102-106}. Several first principle calculations and experimental studies found Mn$_2$NiSn stabilizing into an ordered inverse Heusler structure (Sp.Gr. $F-43m$). \cite{Luo32140634066,Duan.386.102-106,Chieda.486.51-54,Paul110063523,Khatri0000145} 
However, a discrepancy exists related to the experimentally measured and the calculated magnetic moment values. Interestingly, the magnetic moment values calculated by DFT differ wildly from about 0.7 $\mu_{B}$/f.u. to 3.25 $\mu_{B}$/f.u. \cite{Luo32140634066,Paul.116.133903} while the measured net magnetic moment at 5 K for Mn$_2$NiSn is $\sim$ 2.95 $\mu_{B}$/f.u. \cite{Luo32140634066}. In Heusler alloys, the magnetic order is generally determined by the nearest neighbor exchange coupling and the above wide variation in the calculated and the experimental values of magnetic moment point to
a lack of understanding of the magnetic exchange interactions in Mn$_2$NiSn. Electronic structure calculations have suggested two possible scenarios. One considers near-quenching of the Mn moments on the tetrahedral sites and thereby giving a magnetic moment value of 3.25 $\mu_{B}$/f.u. which is fairly close to the observed value \cite{Luo32140634066}. The second approach argues for antisite disorder on the tetrahedral sites leading to the diminishing of the antiferromagnetic interactions and thus escalating the net Mn moment \cite{Paul.116.133903,Paul.23.206003}.

The present work seeks to understand the structural interactions responsible for the observed magnetic properties of Mn$_2$NiSn. Starting with Ni$_2$MnSn, the Ni atoms at the tetrahedral sites are systematically replaced by Mn to realise Ni$_{2-x}$Mn$_{1+x}$Sn ($x$ = 0, 0.25, 0.5, 0.75 and 1.0) alloys. As the $L2_1$ unit cell of Ni$_2$MnSn consists of 8 Ni, 4 Mn and 4 Sn atoms, the chosen values of $x$ on average represent an incremental replacement of one Ni atom in each $L2_1$ unit cell. The alloys have been characterized for their crystal structure, magnetic and transport properties. Using extended x-ray absorption fine structure (EXAFS) at the Mn, Ni and Sn K-edges the near neighbour bond distances between the constituent atoms have been determined and have been used to understand the structural and magnetic ground state in the alloys. Further, temper annealing treatment has been used to elucidate the nature of structural interactions in these alloys.

\section{Experimental}
Polycrystalline alloys of the type Ni$_{2-x}$Mn$_{1+x}$Sn ($x$ = 0, 0.25, 0.5, 0.75 and 1.0) were synthesized by arc melting. For this, the constituent elements, Ni, Mn and Sn (purity $\geq$ 99.99\%) were stoichiometrically weighed and arc melted in an argon atmosphere. The ingots were flipped and remelted several times to ensure homogeneity. Later these ingots were encapsulated in evacuated quartz tubes and annealed at 800$^{\circ}$C for 72 hours before quenching in ice-cold water. These alloys are henceforth referred to as RQ alloys. A part of alloys with $x$ = 0.75 and 1.0 were subjected to a different heat treatment referred to as tempering annealing (TA). After annealing at 800$^{\circ}$C for 72 hours, the quartz tubes containing these two compositions were subjected to an additional heat treatment at 427$^\circ$C for 28 hours and furnace cooled.

An Energy Dispersive X-ray (EDX) analysis, carried out using a scanning electron microscope, confirmed the composition of the alloys to be well within $\sim$ 2\% of the targeted stoichiometry. Room temperature x-ray diffraction (XRD) patterns were recorded using Cu K$\alpha$ radiation in the $20^{\circ}\leq 2\theta \leq 100^{\circ}$ angular range. The diffraction patterns were Rietveld refined using the FULLPROF suite \cite{rodriguez1993juan}. Magnetization measurements as a function of temperature (3 K – 390 K) and magnetic field ($\pm$7 T) using a Quantum Design SQUID magnetometer. Temperature-dependent magnetic measurements were recorded during zero-field  cooled (ZFC), field cooled cooling (FCC) and field cooled warming (FCW) cycles. The samples were first cooled in zero applied field from 300 K down to 3 K and then a field of 100 Oe was applied and the data was recorded during warming (ZFC), subsequent cooling (FCC) and warming (FCW) cycles. Resistivity as a function of temperature was measured in the temperature range of 50 K to 350 K using the conventional DC four-probe technique. EXAFS at the Mn, Ni and Sn  K edges were recorded at the P-65 beamline at PETRA-III synchrotron source (DESY, Hamburg, Germany) at 300 K and 100 K. Absorbers were prepared by sprinkling finely ground powder on a scotch tape and stacking such layers to give an absorption jump $\Delta \mu \leq 1$. Ionization chambers filled with appropriate gases were used to record the incident ($I_{0}$) and transmitted ($I_{t}$) photon intensities simultaneously over an energy range from -300 eV to 1000 eV with respect to the absorption edge energy. The extraction of the EXAFS ($\chi(k)$) signal and its fitting to a structural model were carried out using the well-established procedures in the Demeter program \cite{ravel.12.537-541,ravel.8.314-316}.

\section{Results and Discussion}

\begin{figure}[h]
\begin{center}
\includegraphics[width=\linewidth]{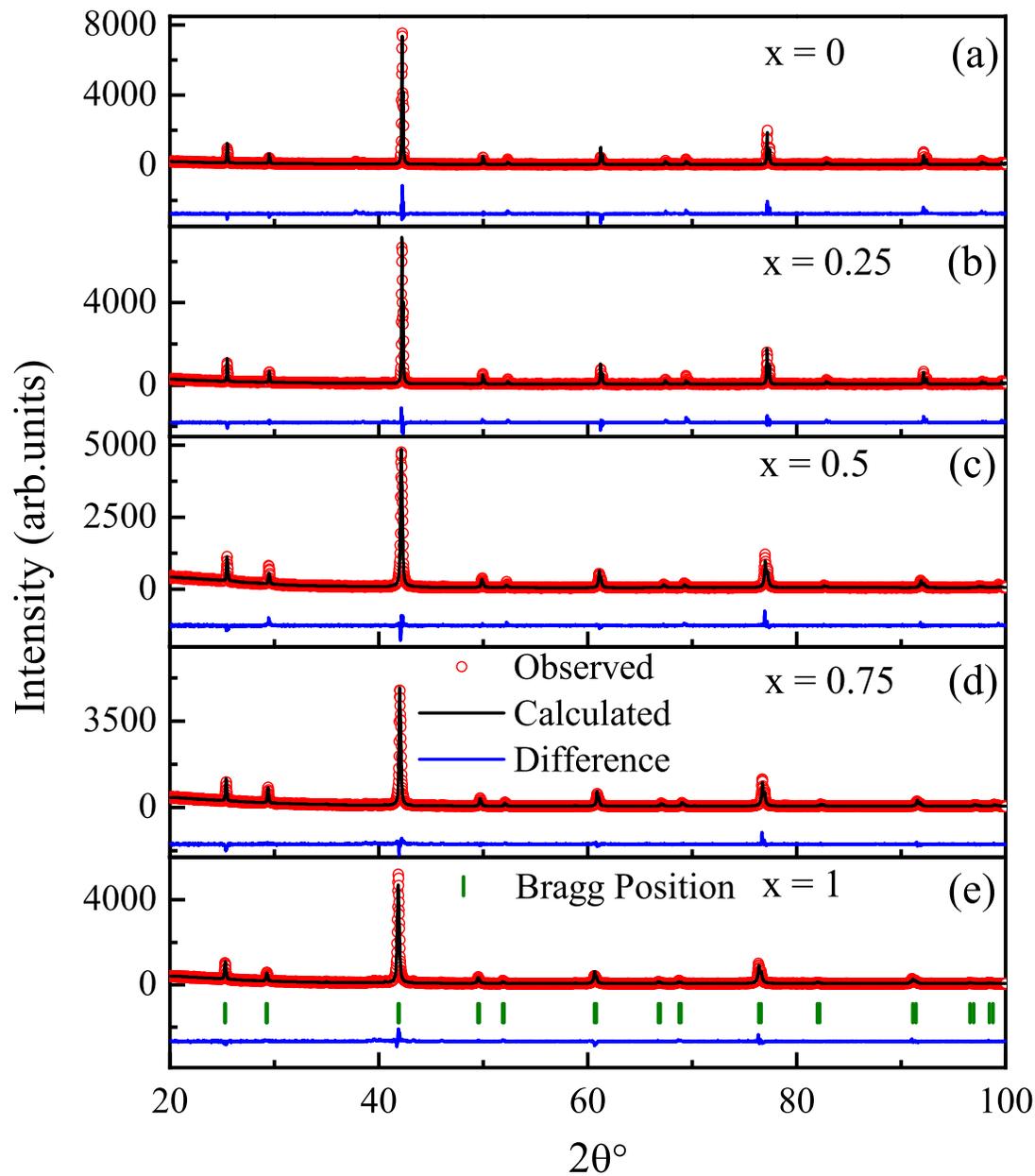}
\caption{Rietveld refined room temperature XRD patterns for Ni$_{2-x}$Mn$_{1+x}$Sn ($x$ = 0, 0.25, 0.5, 0.75 and 1.0) RQ alloys.}
\label{XRD_RQ alloys}
\end{center}
\end{figure}

Rietveld analysed room temperature XRD patterns presented in Fig. \ref{XRD_RQ alloys} confirm the cubic $L2_{1}$ (Space group: $Fm\bar{3}m$) Heusler structure for the RQ series of alloys. The refined values of the lattice constant, $a$, listed in Table.\ref{table_1}, display a monotonic increase with the increase in Mn concentration. Further, the values of $a$ obtained for the two end members, $x=0$ (Ni$_{2}$MnSn) and $x=1$ (NiMn$_{2}$Sn) are in good agreement with those reported in the literature \cite{NAZMUNNAHAR.386.201598,Luo32140634066}.

Temperature-dependent magnetization  M(T), for Ni$_2$MnSn, is presented in Fig. \ref{MT,MH_RQ alloys}a. It displays a ferromagnetic behavior below $T_C  \sim$ 340 K in agreement with literature reports \cite{NAZMUNNAHAR.386.201598}. Similar measurements on other RQ alloys with higher Mn content suggests, the $T_{C}$ systematically increases with increasing Mn substitution (see Table. \ref{table_1}). The $T_C$  could not be determined for compositions with $x \geq 0.5$ as it was well above 390 K. Isothermal magnetisation (M(H)) was recorded by first cooling the sample from room temperature to 3 K in zero applied field and then recording the magnetization as the field was varied in the interval $\pm$ 7 T. Hysteresis loops with low coercive fields presented in Fig. \ref{MT,MH_RQ alloys}b highlight the low-temperature ferromagnetic character of these alloys \cite{NAZMUNNAHAR.386.201598}. Values of the net magnetic moment depicted in the inset of Fig. \ref{MT,MH_RQ alloys}b. were estimated from the M(H) curves by linearly extrapolating the magnetization value at H = 7 T to meet the magnetization axis at H = 0 T. Up to $x$ = 0.5, the magnetic moment remains almost invariant at about 4.0 $\mu_B/f.u.$ and then decreases to 1.93 $\mu_B/f.u.$ in $x$ = 0.75 and again increases to about 2.8 $\mu_B/f.u.$ in $x$ = 1.0 alloy. The sudden change in observed values of magnetic moment hints towards a change in magnetic order for $x \geq 0.75$ alloys.

\begin{figure}
\includegraphics[width=\linewidth]{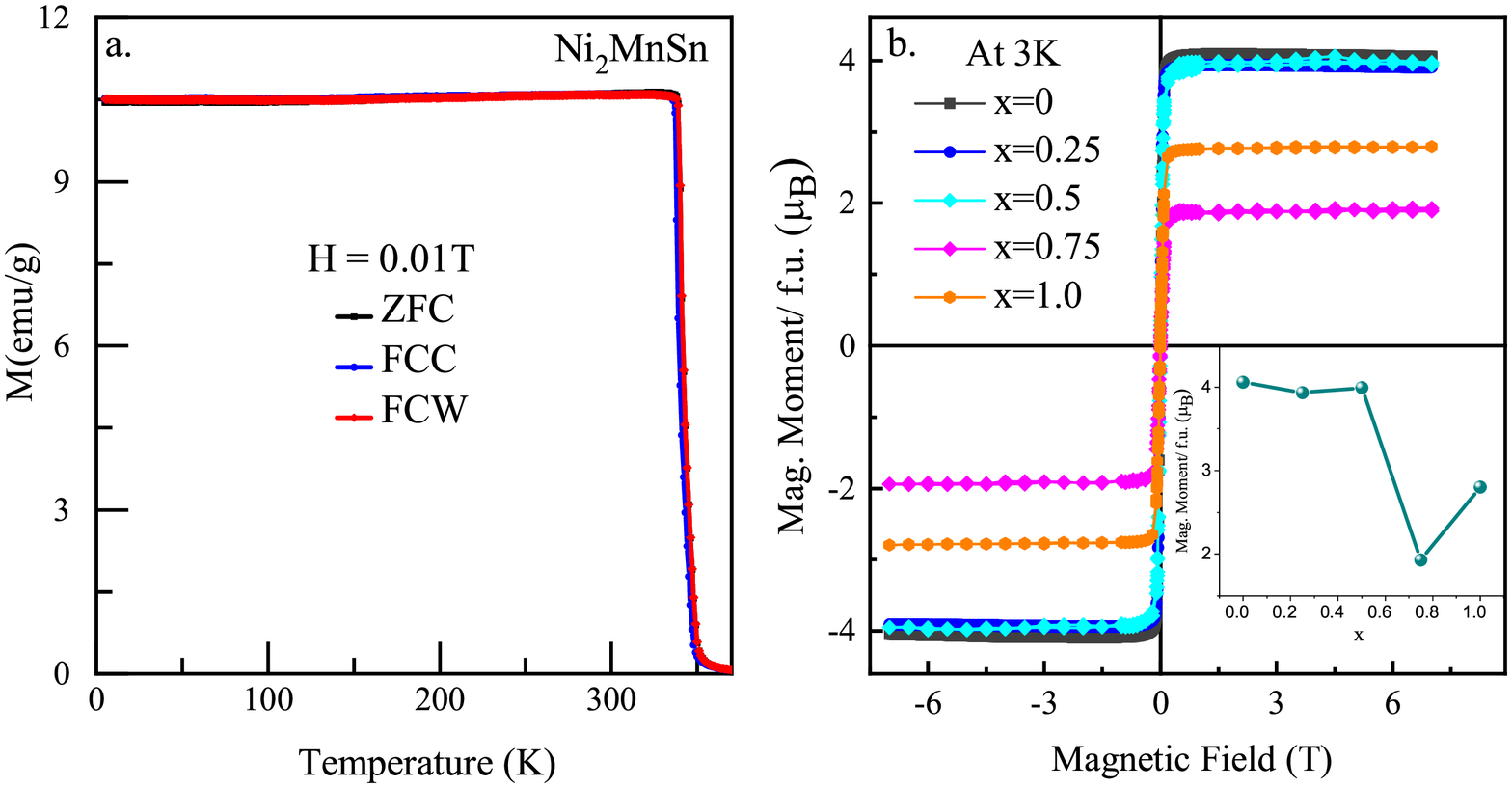}
\caption{a. Temperature dependent magnetization curves recorded for Ni$_2$MnSn in H = 0.01 T during ZFC, FCC, and FCW cycles. b. Isothermal magnetization curves recorded for the Ni$_{2-x}$Mn$_{1+x}$Sn ($x$ = 0, 0.25, 0.5, 0.75 and 1.0) RQ series of alloys at T = 3 K in H = $\pm$7 T.}
\label{MT,MH_RQ alloys}
\end{figure}

\begin{table}[h]
\caption{\label{table_1} Initial and EDX compositions, lattice parameters determined from Rietveld refinements of the XRD patterns, and $T_{C}$ determined from the temperature-dependent magnetization measurements of the quenched Ni$_{2-x}$Mn$_{1+x}$Sn alloys.}
\setlength{\tabcolsep}{0.3pc}
\vspace{0.3cm}
\centering
\begin{tabular}{|c|c|c|c|}
\hline
Alloy & EDX Composition & Lattice parameters (\AA) & T$_{C}$ (K) \\
\hline
Ni$_{2}$MnSn & Ni$_{1.92}$Mn$_{0.99}$Sn$_{1.09}$ & 6.0499 & 340 \\
\hline
Ni$_{1.75}$Mn$_{1.25}$Sn & Ni$_{1.74}$Mn$_{1.22}$Sn$_{1.04}$ & 6.0505 & 358 \\
\hline
Ni$_{1.5}$Mn$_{1.5}$Sn & Ni$_{1.63}$Mn$_{1.53}$Sn$_{0.85}$ & 6.0618 & $>$ 390 \\
\hline
Ni$_{1.25}$Mn$_{1.75}$Sn & Ni$_{1.39}$Mn$_{1.74}$Sn$_{0.86}$ & 6.0803 & $>$ 390\\
\hline
NiMn$_{2}$Sn & Ni$_{1.13}$Mn$_{2.00}$Sn$_{0.87}$ & 6.1027 & $>$ 390\\
\hline
\end{tabular}
\end{table}

\begin{figure}
\includegraphics[width=\linewidth]{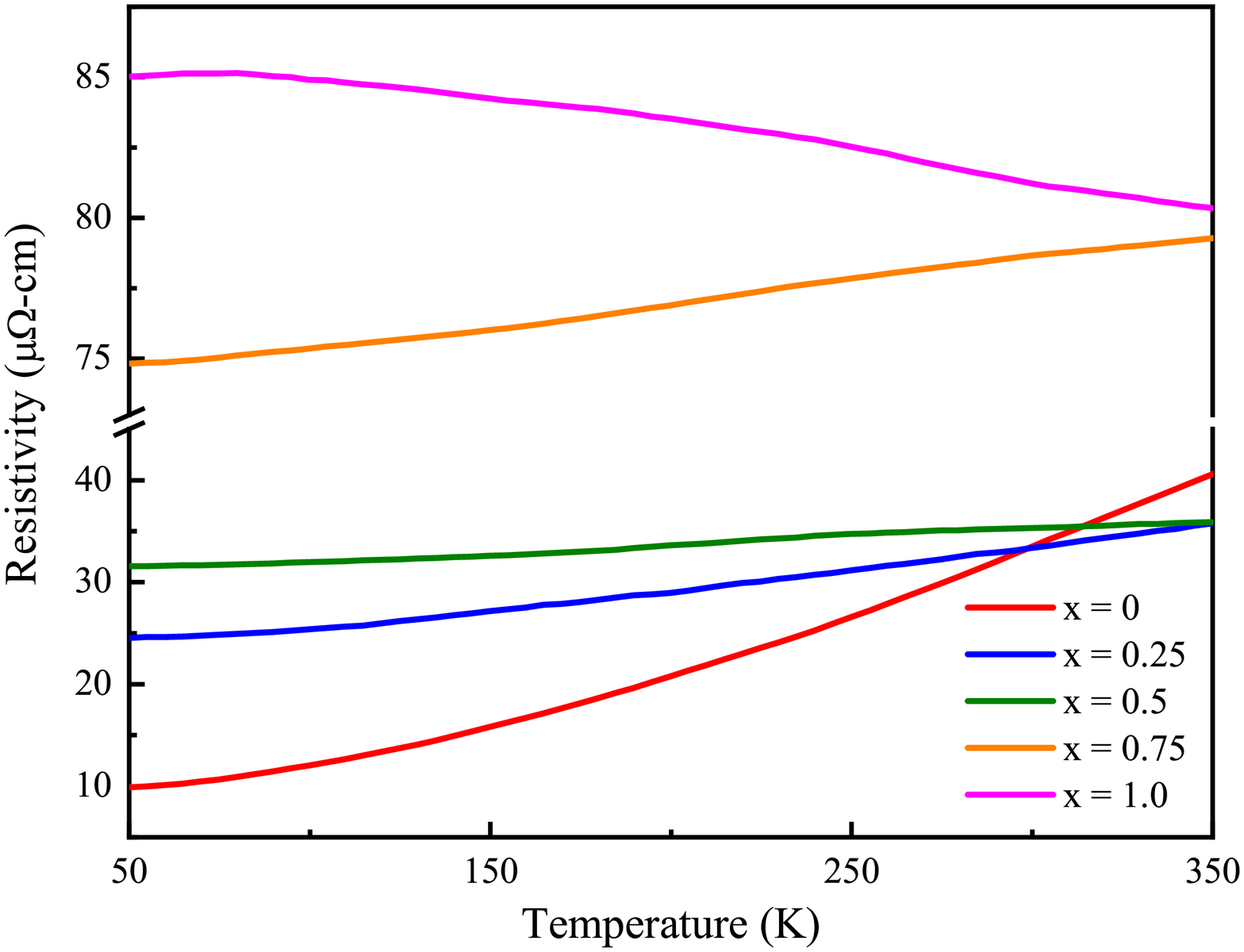}
\caption{Temperature dependent resistivity curves recorded for the Ni$_{2-x}$Mn$_{1+x}$Sn RQ series of alloys.}
\label{Resistivity_RQ alloys}
\end{figure}

Figure \ref{Resistivity_RQ alloys} presents a comparison of the temperature dependence of resistivity ($\rho$(T)) recorded during warming cycles for the RQ series of alloys. While the overall metallic behaviour expected for non-martensitic alloys is preserved for all concentrations up to $x \leq 0.75$, increasing magnitudes of resistivity reflect corresponding changes in the band structure brought about by the replacement of Ni by Mn in the $L2_1$ structure. For $x = 1.0$, the rapidly quenched NiMn$_{2}$Sn alloy exhibits a negative temperature coefficient of resistance across the entire temperature range. The negative $d\rho/dT$ can be ascribed to thermally activated hopping of the conduction electrons trapped in the localized impurity states \cite{ZHANG2021168157}. The impurity states could arise from the formation of structural defects, due to Mn substitution.

\begin{figure}
\includegraphics[width=\linewidth]{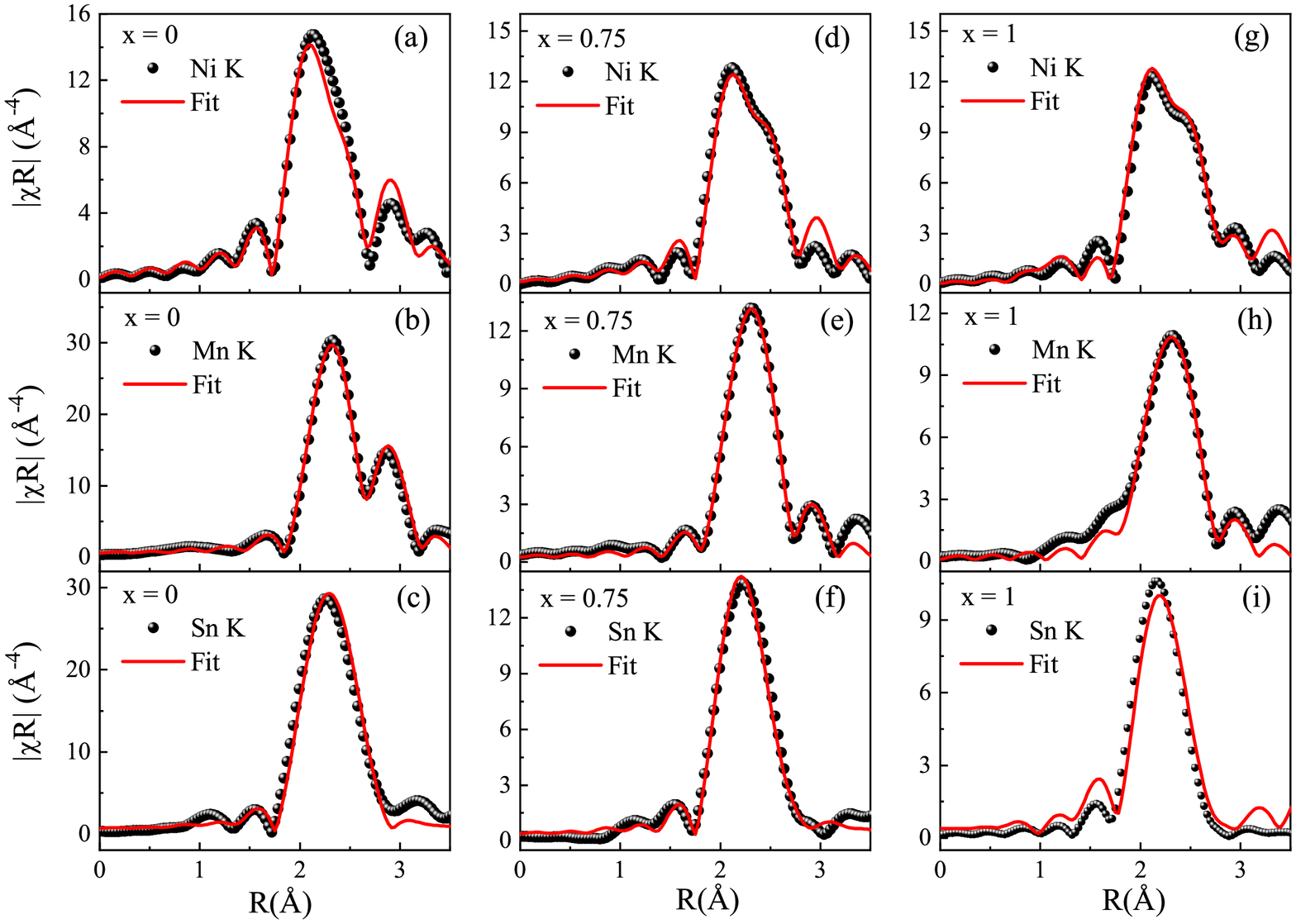}
\caption{Fourier transform magnitudes of the k$^3$ weighted EXAFS at the Ni, Mn and Sn edges along with corresponding best fits in RQ Ni$_{2-x}$Mn$_{1+x}$Sn alloys for $x = 0$ (a-c), $x = 0.75$ (d-f) and $x = 1.0$ (g-i)}
\label{ExafsRQalloys}
\end{figure}

To comprehend the magnetic and transport properties of the RQ alloys, a detailed study of the local structure of the constituent atoms is carried out by analysing RT and LT EXAFS data recorded at Ni, Mn and Sn K edges in Ni$_{2-x}$Mn$_{1+x}$Sn. EXAFS signals at the three K edges were fitted simultaneously in the $k$ range of 3 \AA$^{-1}$ to 12 \AA$^{-1}$) and in the $R$ range of 1.0 \AA~ to 3.0 \AA.  The magnitude of the Fourier transformed (FT) EXAFS data ($|\chi(R)|$) in the $R$ = 1 \AA~ to 3.5 \AA~ range for three selected alloys, ($x$ = 0, 0.75 and 1.0) are presented in Fig. \ref{ExafsRQalloys}.

\begin{table}[h]
\caption{Values of bond distances (R) and mean square disorder in bond distances ($\sigma^2$) obtained by simultaneous fitting of Ni, Mn and Sn EXAFS in Ni$_{2-x}$Mn$_{1+x}$Sn series of RQ alloys, recorded at 100K. Figures in parentheses designate uncertainty in the last digits.}
\label{table2}
\setlength{\tabcolsep}{0.3pc}
\vspace{0.3cm}
\centering
\resizebox{\textwidth}{!}{%
\begin{tabular}{|c|c|c|c|c|c|c|c|c|c|c|}
\hline
\multirow{2}{*}{Bond} & \multicolumn{2}{c|}{Ni$_{2}$MnSn} & \multicolumn{2}{c|}{Ni$_{1.75}$Mn$_{1.25}$Sn} & \multicolumn{2}{c|}{Ni$_{1.5}$Mn$_{1.5}$Sn} & \multicolumn{2}{c|}{Ni$_{1.25}$Mn$_{1.75}$Sn} & \multicolumn{2}{c|}{Mn$_{2}$NiSn} \\ \cline{2-11}
 & R(\AA)& $\sigma^2$(\AA $^2$) & R (\AA) & $\sigma^2$(\AA $^2$) & R(\AA) & $\sigma^2$(\AA $^2$) & R (\AA) & $\sigma^2$(\AA $^2$) & R (\AA) & $\sigma^2$(\AA $^2$) \\ \hline
Ni-Sn & 2.612(1) & 0.0030(2) & \multirow{2}{*}{2.617(3)} & \multirow{2}{*}{0.0031(2)} & \multirow{2}{*}{2.617(6)} & \multirow{2}{*}{0.0040(5)} & 2.617(4) & 0.003(1) & 2.612(8) & 0.004(1) \\ \cline{1-3} \cline{8-11}
MnX-Sn & --- & --- &  &  &  &  & 2.751(16) & 0.008(3) & 2.758(17) & 0.006(2) \\ \hline
Ni-MnY & 2.612(1) & 0.005(1) & \multirow{2}{*}{2.594(3)} & \multirow{2}{*}{0.0048(3)} & \multirow{2}{*}{2.601(8)} & \multirow{2}{*}{2.601(8)} & 2.610(6) & 0.005(1) & 2.596(10) & 0.006(1) \\ \cline{1-3} \cline{8-11}
MnX-MnY & --- & --- &  &  &  &  & \multirow{2}{*}{2.861(56)} & \multirow{2}{*}{0.028(11)} & \multirow{2}{*}{2.818(45)} & \multirow{2}{*}{0.024(11)} \\ \cline{1-7}
MnX-Ni/MnX & --- & --- & \multirow{2}{*}{3.005(6)} & \multirow{2}{*}{0.012(1)} & 2.898(21) & 0.0024(3) &  &  &  &  \\ \cline{1-3} \cline{6-11}
Ni-Ni & 3.016(1) & 0.010(1) &  &  & 3.008(21) & 0.016(3) & 3.004(19) & 0.015(4) & 3.022(57) & 0.021(9) \\ \hline
MnY-Sn & 3.016(1) & 0.006(2) & 2.995(16) & 0.0012(2) & 2.963(20) & 2.963(20) & 2.954(17) & 0.013(2) & 2.943(27) & 0.015(4) \\ \hline
\end{tabular} %
}
\end{table}

In Ni$_2$MnSn with $L2_1$ structure, Ni is surrounded by 4 Mn and 4 Sn atoms in its first coordination shell at about 2.6 \AA. Six Ni atoms at $\sim$ 3.0 \AA~ complete the second shell. On the other hand, Mn and Sn atoms have 8 Ni atoms as their nearest neighbor and 6 Sn or 6 Mn atoms contribute as the second near neighbors respectively. Therefore, a simultaneous fitting of all three, Mn, Ni and Sn EXAFS together results in only four Ni-Mn$_Y$, Ni-Sn, Sn-Mn$_Y$ and Ni-Ni direct scattering correlations instead of seven when these edges are fitted independently. A fitting model employing the structural constraints of the $L2_1$ symmetry with change in bond distance ($\delta R$) and mean square disorder in the bond distance ($\sigma^2$) as variable parameters were conceived. In all the fittings, the coordination numbers were kept fixed to their model values. A similar approach was successfully implemented for fitting the EXAFS data in Ni$_{2+x}$Mn$_{1-x}$Ga alloys \cite{PhysRevB.74.224425}.  The resulting good fit obtained in the $R$ range of 1 \AA~ to 3 \AA, at 50 K is depicted in Fig. \ref{ExafsRQalloys}a and the structural parameters obtained from the fittings are tabulated in Table \ref{table2}. The introduction of Mn in place of Ni to realise in Ni$_{2-x}$Mn$_{1+x}$Sn gives rise to additional three scattering correlations viz, Mn$_X$-Mn$_Y$, Mn$_X$-Sn and Mn$_X$-Ni/Mn$_X$.

In case of the $x = 0.25$ alloy, the contributions of Mn$_X$ cannot be correctly isolated due to lower sensitivity of x-rays in distinguishing between Ni (Z = 28) and Mn (Z = 25) as scatterers at the X sites. However, the  $L2_1$ structural model without the constraints and with individual correlations varying independently resulted in a fairly good fit to the LT EXAFS data at all three edges. Extending this approach to $x = 0.5$ alloy did not result in a good fit and the correlations due to Mn$_X$ atom had to be introduced sequentially. For $x = 0.5$ an Mn-Mn correlation at 2.89 \AA~ was needed to describe the experimental Mn EXAFS well. The observed distance of this correlation is close to the expected Mn$_X$-Ni/Mn$_X$ distance of 3.0 \AA~ and hence we refer to the new Mn-Mn distance as Mn$_X$-Ni/Mn$_X$. With further increase in the value of $x$ to 0.75 and 1.0, another correlation, Mn-Sn had to be introduced to get a good fit to Mn and Sn K-edge EXAFS. However, the best-fitted bond distance of about 2.75 \AA~ was much larger than the expected Mn$_X$-Sn bond length of $\sim$ 2.6 \AA. The Heusler structure of Ni$_{2-x}$Mn$_{1+x}$Sn supports only two Mn-Sn bonds, Mn$_X$-Sn at $\sim$2.6 \AA~ and Mn$_Y$-Sn at about 3.0 \AA. The Mn$_Y$-Sn bond distance is correctly obtained from the fitting and therefore this additional Mn-Sn distance is ascribed to an elongated Mn$_X$-Sn bond. Further, the bond length of Mn$_X$-Ni/Mn$_X$ correlation progressively decreases from 2.9 \AA~ in $x = 0.5$ to 2.8 \AA~ in $x = 1.0$. A similar decrease is also observed in  Mn$_Y$-Sn bond distance as well. The Mn$_Y$-Sn correlation should have a distance of $a/2 \approx $3.025 \AA ($a$ is the lattice parameter). These two, Mn$_X$-Sn and Mn$_X$-Ni/Mn$_X$, misfit bond distances indicate towards a growing local structural disorder around Mn and Sn atoms in Ni$_{2-x}$Mn$_{1+x}$Sn alloys. All the near neighbor distances are listed in Table \ref{table2} and the fits to the Mn, Ni and Sn EXAFS data in $x$ = 0.75 and 1.0 can be respectively seen in Figs. \ref{ExafsRQalloys}(d -- f) and in Figs. \ref{ExafsRQalloys}(g -- i).

\begin{figure}
\includegraphics[width=\linewidth]{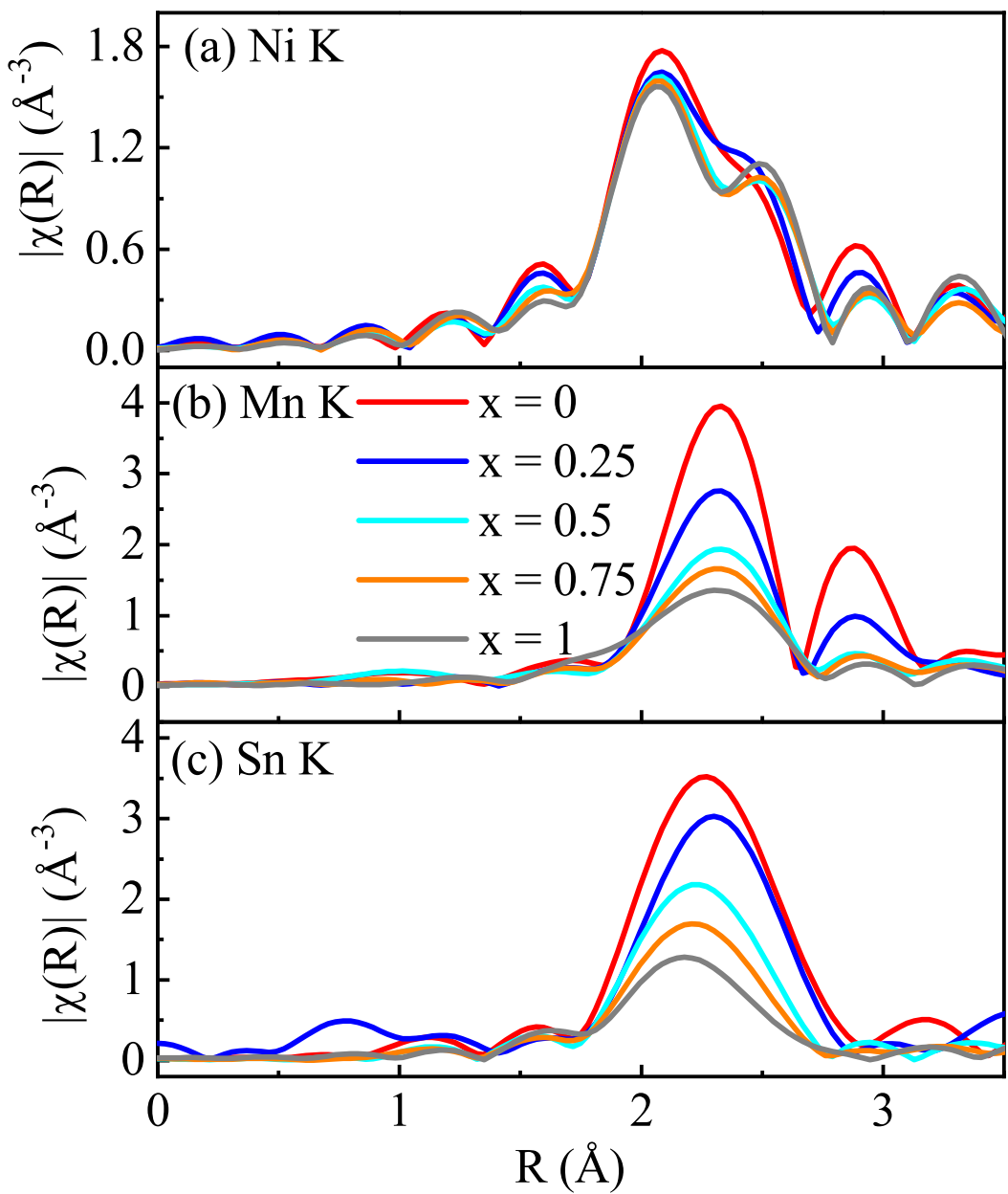}
\caption{A comparison of FT magnitudes of EXAFS data at (a) Ni,(b) Mn and (c) Sn K edges in Ni$_{2-x}$Mn${1+x}$Sn RQ alloys.}
\label{Exafs_comparsion_RQ_alloys}
\end{figure}

A comparison of crystal structures of several alloys containing Mn and Sn suggests that the observed Mn-Sn bond distance is very close to the bond distances reported in Mn$_{3}$Sn type hexagonal $D0_{19}$ structure. Mn$_3$Sn type environment is possible in such Ni$_{2-x}$Mn$_{1+x}$Sn with an antisite disorder among the Mn and Sn sublattices of the Heusler alloy. To check the possibility of such a disorder, the magnitudes of FT of EXAFS as the Mn, Ni and Sn K edges are compared in Fig.\ref{Exafs_comparsion_RQ_alloys}. Little or no variation of FT amplitudes observed for Ni K edge (Fig.\ref{Exafs_comparsion_RQ_alloys}a) suggests very little change in the local structural environment of Ni with Mn substitution in Ni$_2$MnSn. On the other hand, a systematic decrease in scattering amplitudes with increasing Mn concentration observed for the Mn and Sn K edges in Figs.\ref{Exafs_comparsion_RQ_alloys}b. and c. indicates the presence of disorder around the Mn and Sn atoms. Such structural defects are responsible for phase separation and the phenomenon of shell ferromagnetism in the Ni$_{2}$Mn$_{2-x}$In$_{x}$ alloys \cite{PhysRevB.104.054101}. Upon temper annealing, these alloys phase separate into proportionate amounts of Heusler (Ni$_{50}$Mn$_{25}$In$_{25}$) and $L1_0$  (Ni$_{50}$Mn$_{50}$) phases \cite{Caklr2016}. On similar lines, the Ni$_{2-x}$Mn$_{1+x}$Sn alloys can segregate into Ni$_{2}$MnSn and Mn$_{3}$Sn type phases. Segregation of Mn$_3$Sn type defects in Ni$_{2-x}$Mn$_{1+x}$Sn alloys would require a presence of $B2$ type antisite disorder in them. Using Webster's approach \cite{webster1969}, an attempt was made to calculate the possible $B2$ type disorder in all as prepared compositions of Ni$_{2-x}$Mn$_{1+x}$Sn alloys. A maximum of 11\% of $B2$ disorder was obtained in NiMn$_2$Sn and it progressively decreased to negligible value in Ni$_{1.75}$Mn$_{1.25}$Sn.

\begin{figure}
\includegraphics[width=\linewidth]{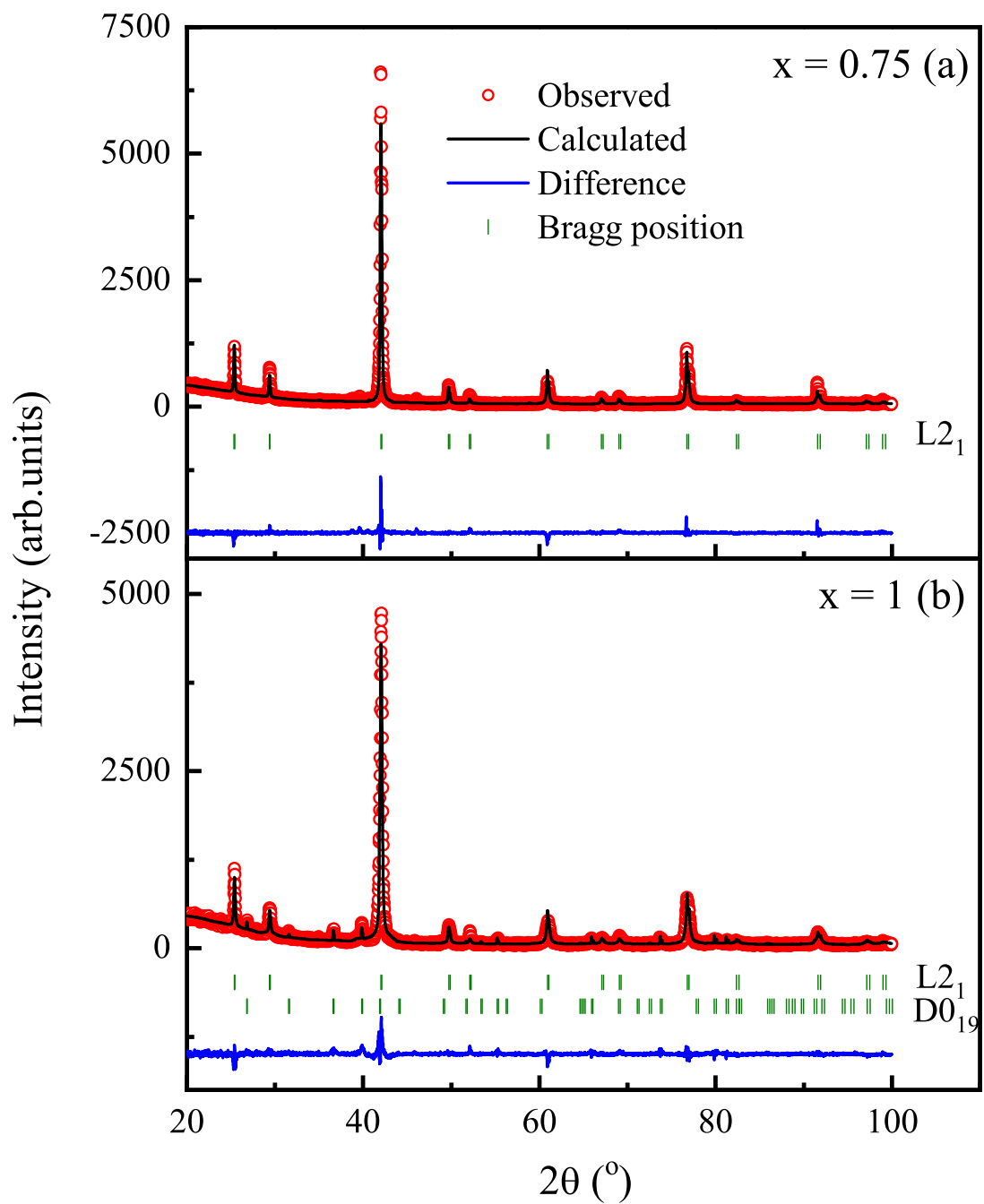}
\caption{Rietveld refined XRD patterns for (a) Ni$_{1.25}$Mn$_{1.75}$Sn and (b) NiMn$_{2}$Sn TA alloys}
\label{XRD_TA_alloy}
\end{figure}

To explore the possibility of phase separation in Mn-rich Ni$_{2-x}$Mn$_{1+x}$Sn alloys, the compositions, $x$ = 0.75 and 1.0 were temper annealed and investigated for their structure, magnetic and transport properties. Rietveld refined x-ray diffraction patterns recorded for the two temper annealed alloys ($x$ = 0.75 and 1.0) are presented in Fig. \ref{XRD_TA_alloy}. The diffraction pattern of $x = 1.0$ TA alloy consists of two phases, a major L2$_{1}$ ($\sim$ 93\%) and a minor ($\sim$ 7\%) Mn$_{3}$Sn type hexagonal phase (Space group: P6$_{3}$/mmc). The $x$ = 0.75 however, presents itself as a single-phase alloy with cubic L2$_{1}$ Heusler structure. Studies on Ni$_{50}$Mn$_{50-x}$Z$_x$ have shown that the extent of phase separation depends on the time of annealing \cite{Dincklage2018}. Further, lower the defect concentration, more annealing time is required for phase separation. Therefore, in the annealing time used here, the alloy with a larger concentration of structural defects (NiMn$_2$Sn) exhibits two crystallographic phases while Ni$_{1.25}$Mn$_{1.75}$Sn presents a single-phase structure.

The magnetic and transport properties of these temper annealed alloys do show considerable changes in comparison to their RQ counterparts. The magnetic isotherms recorded at 3 K in a field of $\pm$ 7 T for the two alloy compositions are presented in Fig.\ref{MH,Resistivity_TA alloys}a. While the net magnetic moment in the phase-separated Mn$_{2}$NiSn  alloy increases marginally to 2.83 $\mu_B$/f.u., the magnetic moment in Ni$_{1.25}$Mn$_{1.75}$Sn TA alloy displays a significant increase from 1.93 $\mu_B$/f.u. in RQ alloy to  3.31 $\mu_B$/f.u. in TA alloy. Further, the resistivity as a function of temperature in Fig.\ref{MH,Resistivity_TA alloys}b exhibits a decrease in magnitude for both the TA alloys as compared to the corresponding RQ alloys and the temperature coefficient of resistance changes sign from negative to positive  in NiMn$_2$Sn TA alloy.

\begin{figure}
\includegraphics[width=\linewidth]{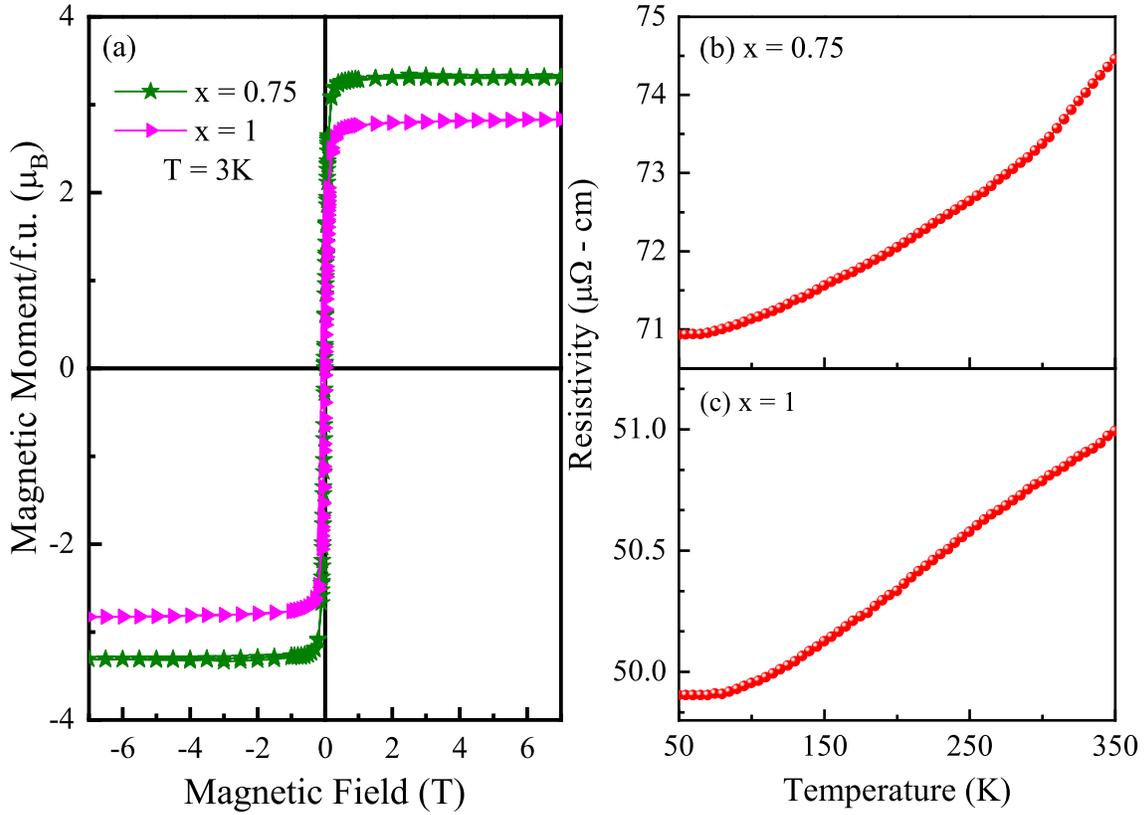}
\caption{(a) Magnetic isotherms at 3K in $\pm$7 T, (b) Temperature dependent resistivity for Ni$_{1.25}$Mn$_{1.75}$Sn and NiMn$_{2}$Sn TA alloys.}
\label{MH,Resistivity_TA alloys}
\end{figure}

\begin{figure}
\includegraphics[width=1\linewidth]{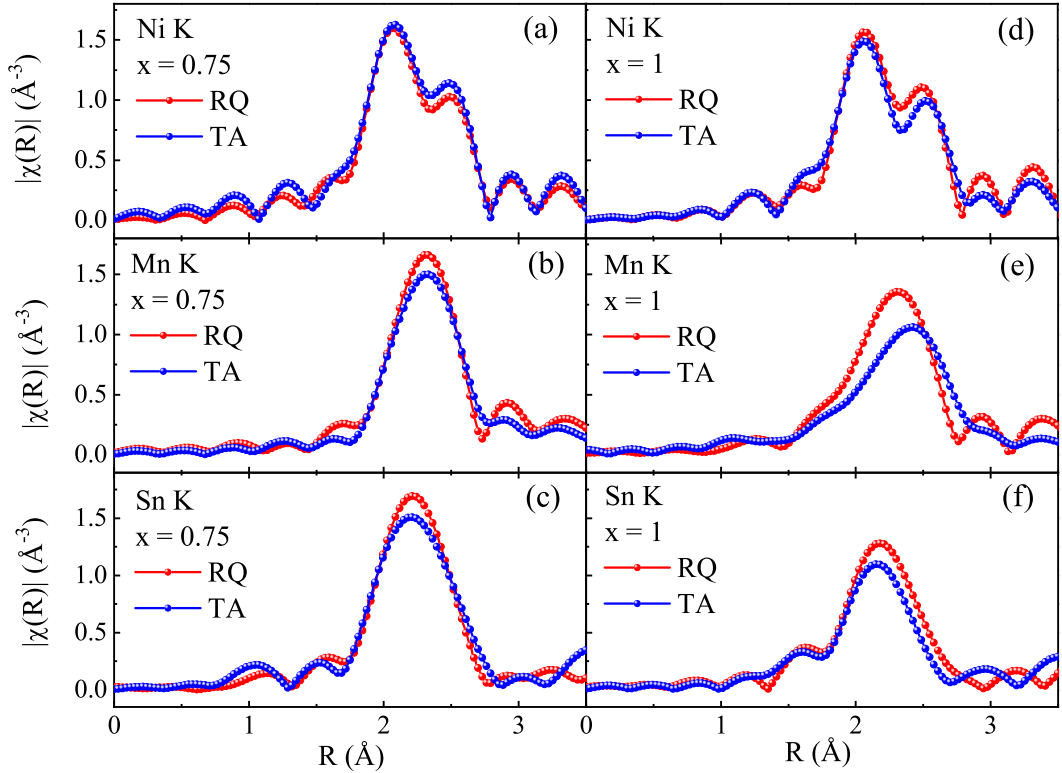}
\caption{A comparison of FT magnitude of the Ni, Mn and Sn EXAFS in RQ and TA Ni$_{1.25}$Mn$_{1.75}$Sn and NiMn$_2$Sn alloys.}
\label{ExafsRQTAcomparison}
\end{figure}

Recent EXAFS studies on temper annealed Ni$_2$Mn$_{2-x}$In$_x$ alloys have shown that complete phase separation is present at least at the local structural level even if the alloy presents a single-phase structure in XRD \cite{PhysRevB.104.054101}. Therefore, to check the extent of phase separation and the changes in the local structural environment due to temper annealing, the LT EXAFS data recorded at the Mn, Ni and Sn K edges in Ni$_{1.25}$Mn$_{1.75}$Sn and NiMn$_2$Sn TA alloys are compared with the respective EXAFS data recorded in the RQ alloys (see Fig. \ref{ExafsRQTAcomparison}). The comparison shows clear deviations in the FT magnitude of Mn and Sn EXAFS in $x$ = 1.0 while subtle deviations are also noted for $x$ = 0.75 alloys. In both the alloys, the Ni local structure remains relatively unaffected due to temper annealing, thus giving weight to the possibility of phase separation of these Mn-rich Ni$_{2-x}$Mn$_{1+x}$Sn TA alloys into Ni$_2$MnSn type Heusler and Mn$_3$Sn type $D0_{19}$ phases.

\begin{figure}
\includegraphics[width=\linewidth]{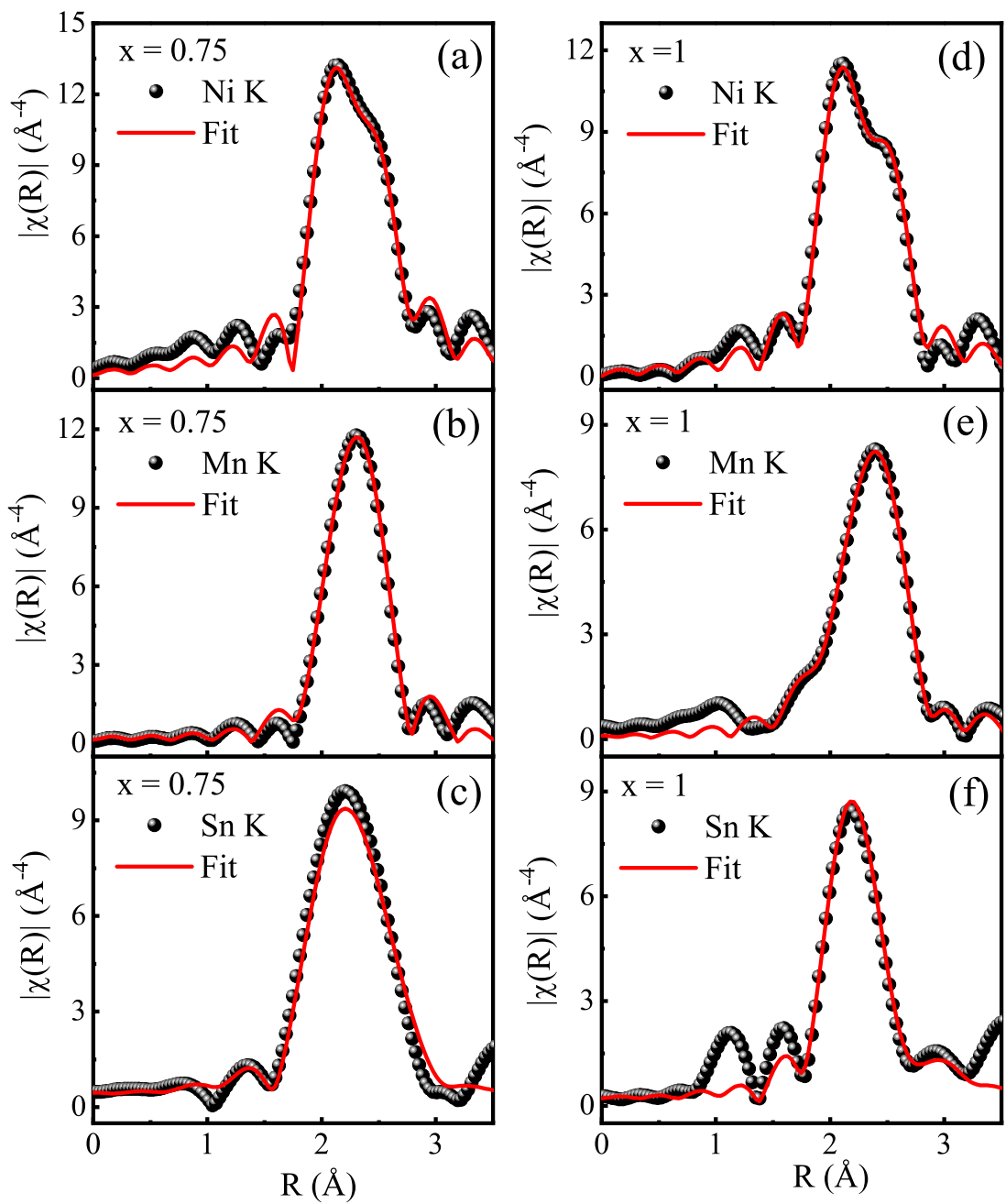}
\caption{FT magnitude of $k^3$ weighted EXAFS at Ni, Mn and Sn K edges and the corresponding best fits in Ni$_{2+x}$Mn$_{1-x}$Sn ($x$ = 0.75, 1) TA alloys.}
\label{Exafs_TA_alloy}
\end{figure}

To further confirm the phase separation of the TA alloys, EXAFS data at all three edges were fitted to a phase-separated model consisting of near neighbor  correlations from Ni$_2$MnSn and Mn$_3$Sn structures. The resulting good fits are depicted in Fig. \ref{Exafs_TA_alloy} and the corresponding best-fit parameters are summarised in Table.\ref{table_3}. The fitted percentage phase fractions in the TA alloys indicate a complete phase separation of the two TA alloy compositions into $L2_1$ and $D0_{19}$ phases. However, the two Mn-Mn and Mn-Sn correlations of the hexagonal phase could not be individually fitted in Ni$_{1.25}$Mn$_{1.75}$Sn. They were combined and considered as one Mn-Mn and one Mn-Sn correlation. In NiMn$_2$Sn, the two Mn-Mn and two Mn-Sn correlations were fitted individually as can be seen in Table \ref{table_3}.

\begin{table}[h]
\caption{Bond distances (R) and mean square disorder in bond distance ($\sigma^2$) obtained by simultaneous fitting of Ni, Mn and Sn EXAFS recorded at 100K in Ni$_{2+x}$Mn$_{1-x}$Sn ($x$ = 0.75, 1) TA alloys. Figures in parentheses designate uncertainty in the last digit.}
\label{table_3}
\setlength{\tabcolsep}{0.3pc}
\vspace{0.3cm}
\centering
\begin{tabular}{|c|c|c|c|c|c|c|}
\hline
\multirow{2}{*}{Bond} & \multicolumn{3}{c|}{Ni$_{1.25}$Mn$_{1.75}$Sn} & \multicolumn{3}{c|}{Mn$_{2}$NiSn} \\ \cline{2-7}
 & \begin{tabular}[c]{@{}c@{}}Phase \\ Fraction \%\end{tabular} & R (\AA) & $\sigma^2$(\AA $^2$) & \begin{tabular}[c]{@{}c@{}}Phase \\ fraction \%\end{tabular} & R (\AA) & $\sigma^2$(\AA $^2$) \\ \hline
Ni-Sn & \multirow{4}{*}{0.63(5)} & 2.616(4) & 0.004(1) & \multirow{4}{*}{0.50(4)} & 2.615(5) & 0.006(1) \\ \cline{1-1} \cline{3-4} \cline{6-7}
Ni-Mn &  & 2.592(6) & 0.006(1) &  & 2.580(5) & 0.006(1) \\ \cline{1-1} \cline{3-4} \cline{6-7}
Ni-Ni &  & 3.026(26) & 0.021(4) &  & 3.021(15) & 0.021(4) \\ \cline{1-1} \cline{3-4} \cline{6-7}
Mn-Sn &  & 3.045(40) & 0.009(3) &  & 2.994(19) & 0.004(1) \\ \hline
Mn-Mn & \multirow{4}{*}{0.37(5)} & \multirow{2}{*}{3.140(27)} & \multirow{2}{*}{0.010(4)} & \multirow{4}{*}{0.50(4)} & 2.723(7) & 0.005(1) \\ \cline{1-1} \cline{6-7}
Mn-Mn &  &  &  &  & 2.901(17) & 0.007(2) \\ \cline{1-1} \cline{3-4} \cline{6-7}
Mn-Sn &  & \multirow{2}{*}{2.770(19)} & \multirow{2}{*}{0.011(3)} &  & 2.725(12) & 0.004(1) \\ \cline{1-1} \cline{6-7}
Mn-Sn &  &  &  &  & 3.102(32) & 0.011(4) \\ \hline
\end{tabular}
\end{table}

The phase-separated model was also applied to the two RQ alloy compositions with $x$ = 0.75 and 1.0. Similar good fits were obtained for both the alloys but the phase fractions of the two phases were different. In the case of Ni$_{1.25}$Mn$_{1.75}$Sn the fractions of the $L2_1$ and $D0_{19}$ phases respectively were $65 \pm 4 \%$ and $35 \pm 4 \%$ and in NiMn$_2$Sn the percentage fractions of the cubic and hexagonal phases were obtained to be $73 \pm 16 \%$  and $27 \pm 16 \%$ respectively. This observation of phase separation at the local structural level confirms the formation of a defect phase within the overall $L2_1$ symmetry. The near neighbour Mn-Mn and Mn-Sn bond distances in Mn$_3$Sn are much longer than the Mn-Mn and Mn-Sn bond distances in the $L2_1$ phase. Therefore, the formation of such elongated bonds within the $L2_1$ symmetry should exert pressure on other near neighbour distances. Indeed such a contraction of Mn$_X$-Ni/Mn$_X$ and Mn$_Y$-Sn bonds is observed in Ni$_{2-x}$Mn$_{1+x}$Sn RQ alloys with $x \ge 0.5$.

The formation of the structural defects and the eventual phase separation on temper annealing also explains the observed magnetic and transport properties. The high value of resistance and the negative temperature coefficient of resistance in rapidly quenched NiMn$_2$Sn is a result of the scattering of charge carriers by the structural defects. Temper annealing leads to an overall decrease in resistance and a change in character in the case of NiMn$_2$Sn. This can be ascribed to the metallic nature of the two phases formed upon temper annealing. Similarly, the increasing fraction of ferromagnetic Ni$_2$MnSn type Heusler component from 50\% in NiMn$_2$Sn to 63\% in Ni$_{1.25}$Mn$_{1.75}$Sn explains the increase in the magnetic moment from 2.83 $\mu_B$/f.u. to 3.31 $\mu_B$/f.u.

\section{Conclusion}

In conclusion, the Mn substitution in Ni$_{2-x}$Mn$_{1+x}$Sn leads to a formation of Mn$_3$Sn type structural defects. These defects are initiated by an antisite disorder between Sn occupying the Z sublattice and Mn at the Y sublattice. These structural defects segregate and phase separate upon temper annealing. Defect concentration seems to increase with the increase in Mn substitution at the X site and thus this substituted Mn could be the precursor for Y-Z antisite disorder. The structural defects also explain the observed magnetic and transport properties.

\section*{Acknowledgements}
The authors would like to acknowledge Science and Engineering Research Board (SERB) for the financial assistance under project EMR/2017/001437 and the travel support from the Department of Science and Technology, Govt. of India within the framework of India DESY collaboration. We thank DESY (Hamburg, Germany), a member of Helmholtz Association HGF, for the provision of experimental facilities. Parts of this research were carried out at PETRA III and we thank Edmund Welter and Ruidy Nemausat for experimental assistance. Beamtime was allocated for the proposal (I-20180679). Some of the EXAFS experiments were also performed at BL-12C at Photon Factory under the proposal No. 20160676 and the help received is also acknowledged.

\bibliographystyle{iopart-num}
\bibliography{mybibfile}

\end{document}